\documentclass[pdflatex,sn-nature,iicol]{sn-jnl}

\usepackage{graphicx}
\usepackage{amsmath,amssymb,amsfonts}
\usepackage{booktabs}

\theoremstyle{thmstyleone}

\begin{document}

\title[Goldene to Noblene]{From Goldene to Noblene: exhaustive enumeration of the ordered Au-Ag-Cu monolayer alloys}

\author[1,2]{\fnm{Marcelo} \sur{Lopes Pereira Junior}}\email{marcelo.lopes@unb.br}

\affil[1]{\orgdiv{University of Bras\'ilia},
\orgname{College of Technology, Department of Electrical Engineering}, \orgaddress{\postcode{70910-900}, \city{Bras\'ilia}, \state{Federal District}, \country{Brazil}}}

\affil[2]{\orgdiv{NanoEngineering Laboratory},
\orgname{College of Technology, University of Bras\'ilia}, \orgaddress{\postcode{70910-900}, \city{Bras\'ilia}, \state{Federal District}, \country{Brazil}}}

\abstract{Two-dimensional metals became a laboratory reality with the isolation of Goldene, a gold sheet one atom thick released by chemical exfoliation, which raises the question of what alloying can achieve on the same close-packed lattice. Here we combine an exhaustive enumeration of derivative superstructures with density functional theory to map the ordered Au-Ag-Cu monolayer alloys, relaxing every symmetry-inequivalent arrangement up to four atoms per cell and extending the set with larger cells selected by a fitted model. We find that the arrangement of the atoms, and not their proportion, controls the mixing energy, since the spread among the orderings of a single composition is several times the step between neighboring compositions, one composition spans more than 50~meV/atom between two arrangements of the same three atoms, and at two compositions the arrangement decides the sign. The energetics is carried by a competition between Au-Cu contacts, which bind, and Ag-Cu contacts, which do not, and bulk enthalpies computed under the same protocol show that this chemistry is not uniformly rescaled in two dimensions, since Au-Cu orders more strongly in the monolayer than in the crystal while the other two weaken. A pair model completed by an elastic size-mismatch term reproduces the computed energies to a few meV/atom and returns a composition landscape whose convex hull no balanced ternary composition reaches. One equimolar ordering, which we name Noblene, is the only structure in which every atom is surrounded exclusively by unlike species, and it is dynamically stable and metallic, and carries a Poisson's ratio above one half, higher than that of any of the three pure monolayers. Since Noblene sits on the lattice that Goldene already realizes, the route that produced Goldene is a plausible starting point for its synthesis.}


\maketitle

\section{Introduction}\label{sec:intro}

The isolation of atomically thin crystals established a family of two-dimensional (2D) materials whose properties are set by a single atomic plane rather than by a bulk lattice~\cite{Novoselov2005}, and its metallic members are those for which the reduction of thickness carries the most direct technological consequence. In a metal one atom thick, every atom is a surface atom, so the fraction of the material accessible to an adsorbate reaches its upper bound and the mass-specific activity is maximized, which is what has driven the preparation of metallenes for electrocatalysis and for the wider set of surface-controlled functions~\cite{Prabhu2021, Yu2023}. Confining the conduction electrons to a single plane further modifies the electronic screening, the plasmonic response, and the charge transport of a metal that would otherwise behave as a three-dimensional conductor, so that a 2D metal is a distinct material rather than a thinner version of its parent~\cite{Yu2023, Kashiwaya2025}. The interest in these layers, which spans catalysis, sensing, plasmonics, and ultrathin interconnects, therefore rests on structures that are considerably more difficult to obtain than the layered compounds from which the field began.

Metallic monolayers were predicted by electronic structure calculations long before they were realized experimentally, because metallic bonding is nondirectional and provides no intrinsic mechanism to terminate growth at a single atomic layer. A survey of the elemental 2D lattices across the periodic table nevertheless identified the close-packed hexagonal sheet as the most stable geometry available to this family, retaining about 70\% of the bulk cohesive energy even after the coordination is halved from twelve neighbors to six~\cite{Nevalaita2018}. That prediction was confirmed with the isolation of Goldene from a MAX phase, one of the layered ternary carbides and nitrides of the M$_{n+1}$AX$_n$ family, in which gold occupies the site normally held by the main-group element. The surrounding titanium carbide slabs were dissolved by wet chemical etching, and the gold was recovered as a free-standing sheet one atom thick~\cite{Kashiwaya2024}.

The route by which the layer is obtained constrains the degrees of freedom subsequently available to it, since a sheet released from a parent crystal rather than grown on a support inherits the close-packed arrangement of the site it vacated and is therefore subject to no epitaxial constraint. In this sense, the lattice is fixed by the precursor, and the composition of the layer remains the only variable open to manipulation. First-principles descriptions converge on that geometry, reporting a hexagonal $P6/mmm$ cell holding a single atom with six in-plane neighbors~\cite{Sheremetyeva2025}, an in-plane elastic response that is isotropic~\cite{Mortazavi2024}, and a lattice that is essentially equilateral and triangular~\cite{Wang2025}. Stacked assemblies built from the same layer also retain the in-plane motif~\cite{Pereira2025}, which indicates that the close-packed sheet is a robust structural unit rather than a geometry confined to the monolayer limit.

The family has expanded rapidly around that layer, since the same exfoliation strategy has been extended to a trilayer obtained from a different parent phase~\cite{Shi2025}, an independent confinement route has delivered a wider set of metals at the angstrom thickness limit~\cite{Zhao2025}, and noble metallenes now constitute a defined class with established synthesis routes and application prospects~\cite{Kashiwaya2025}. The two lighter group-11 metals have been placed on the Goldene lattice and are dynamically stable there, with elastic, electronic, and optical signatures distinct from gold and from each other~\cite{dosSantos2025, Wang2025, Xu2026}. Neither has yet been isolated, and Silverene and Copperene therefore serve as reference systems against which any group-11 monolayer is assessed. Other motifs have been examined for the same three metals, including threefold-coordinated honeycombs~\cite{Kapoor2018} and kagome networks~\cite{Bastos2026}, although the close-packed sheet remains the geometry available from current synthesis.

The availability of three elements on a common lattice makes alloying the natural next step, and the group-11 triad is the appropriate ternary to examine first, since its bulk phase behavior spans the entire range of mixing tendencies while the chemistry remains within a single column of the periodic table. Silver and gold form a continuous solid solution~\cite{Okamoto1983}, gold and copper order into the L1$_0$ and L1$_2$ superstructures at low temperature~\cite{Okamoto1987}, and silver and copper are only sparingly miscible, separating across a wide miscibility gap~\cite{Subramanian1993}. Ozoli\c{n}\v{s} \textit{et al.} reconciled the first two cases, showing that Ag-Au does form ordered compounds with a negative mixing enthalpy, but with ordering temperatures low enough that the disordered solution is what experiment observes~\cite{Ozolins1998}. A thermodynamic assessment of the full ternary reproduces the same hierarchy through a positive Cu-Ag interaction parameter set against negative Cu-Au and Ag-Au ones~\cite{Cao2007}. Strong ordering, weak ordering, and phase separation are therefore all accessible within a single chemical system, which makes the group-11 triad an unusually informative test case for the effect of dimensionality on alloy thermodynamics.

How that chemistry behaves once the metal is reduced to a single atomic layer has not been established, since the only first-principles study of alloyed noble-metal monolayers treats the Au-Ag and Au-Cu binaries at equal proportions, without competing orderings at fixed composition, without the Ag-Cu pair, and without the ternary~\cite{Wang2025b}. That study further predicts that alloying drives the layer off the equilateral lattice toward an isosceles one, a prediction that systematic enumeration can test, while reviews of the field identify single-atom-thick alloy monolayers as an open direction~\cite{Kashiwaya2025}. The enumeration itself has therefore not been carried out, although a lattice of sixfold coordination is precisely where the distinction between one arrangement and another can be expected to be sharpest.

\begin{figure*}[t]
\centering
\includegraphics[width=0.8\linewidth]{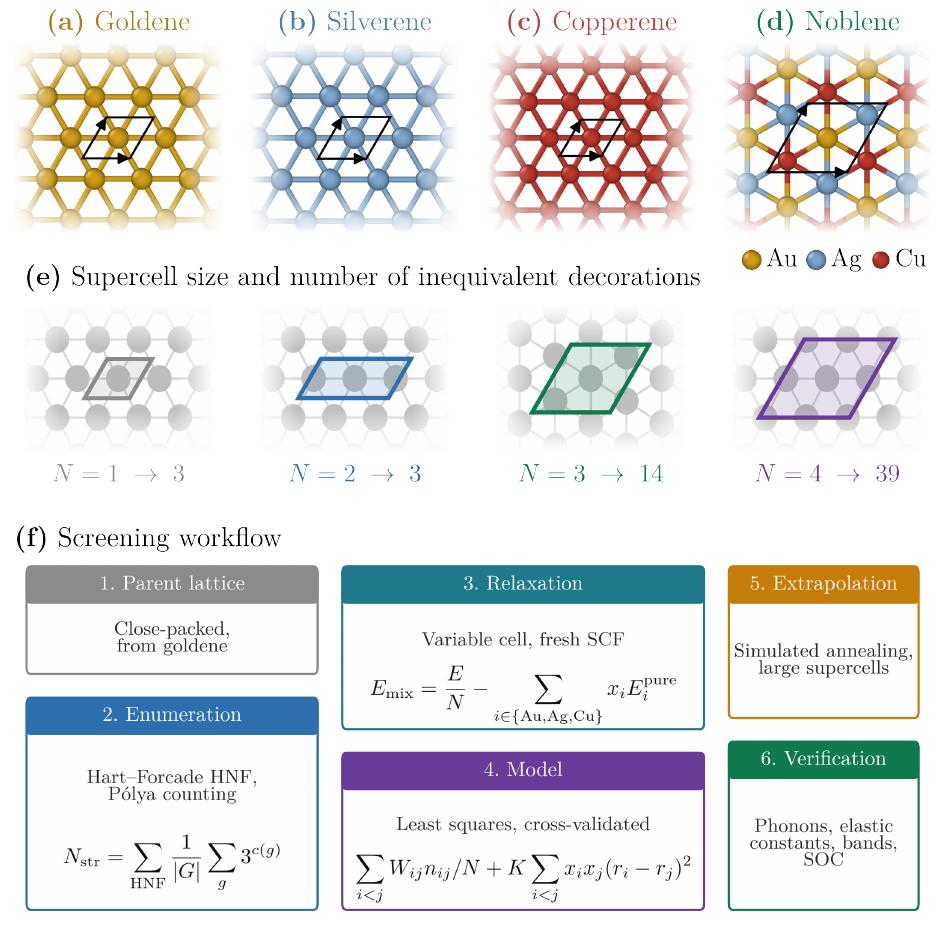}
\caption{The four reference monolayers on the close-packed lattice inherited from Goldene, Goldene (a), Silverene (b), Copperene (c), and Noblene (d). The same lattice with cells containing one through four atoms and the number of symmetry-inequivalent decorations of each (e), and the enumeration and screening workflow (f).}
\label{fig:fig01}
\end{figure*}

In this work, we have carried out that enumeration, generating the derivative superstructures of the Goldene lattice and relaxing them within density functional theory (DFT). The enumeration is exhaustive through four atoms per cell. We have extended it to larger cells with a model fitted to the exhaustive set, and have examined the resulting structures for their energetics, phonon spectra, electronic structure, and elastic response. Figure~\ref{fig:fig01} presents the four reference monolayers, the supercells enumerated at each size, and the workflow through which the set was built and screened. We find that ordering dominates composition in setting the mixing energy, that the balance between Au-Cu and Ag-Cu contacts is what sets it, and that a single equimolar arrangement, which we name Noblene, stands apart because no atom in it has a neighbor of its own kind.

\section{Results and discussion}\label{sec:results}

The results below follow the screening workflow of Figure~\ref{fig:fig01}. Its first three stages, the parent lattice, the enumeration, and the relaxation that fixes the energy scale, are taken up first, followed in turn by the ordering degree of freedom, by the model that carries the enumeration beyond the cells that direct calculation can reach, and by the verification of the structure that the enumeration singles out.

\subsection{Systems and enumeration}\label{sec:systems}

Figure~\ref{fig:fig01}(a) through (d) shows the four monolayers that serve as the reference systems of this work, the three pure ones against which every mixing energy is measured and the ordered equimolar ternary that Section~\ref{sec:Noblene} characterizes. Goldene, Silverene, and Copperene are the single close-packed planes of the three group-11 metals, each a hexagonal $P6/mmm$ cell holding one atom whose six in-plane neighbors are equidistant, so that the plane is a triangular lattice. Noblene is the equimolar ternary ordering built on that same plane, in which Au, Ag, and Cu occupy the three sublattices of a cell three times larger than the parent one and rotated with respect to it. That cell is the $\sqrt{3}\times\sqrt{3}$~R30$^{\circ}$ supercell, whose lattice vectors are $\sqrt{3}$ times longer than those of the parent cell and rotated by 30$^{\circ}$ from them, and it is the smallest cell of the close-packed lattice on which the three species can be placed on an equal footing. We refer to the arrangement it carries as Noblene throughout, and the supercell notation is not repeated once the structure has been introduced. Noblene relaxes in the space group $P\bar{6}m2$, so that alloying lowers the symmetry from $P6/mmm$, but the threefold rotation of the parent lattice survives, and in two dimensions that rotation alone is enough to make the in-plane elastic response isotropic, as Section~\ref{sec:Noblene} confirms by direct calculation.

The nearest-neighbor distances of the pure monolayers are 2.739, 2.789, and 2.416~\r{A} for Au, Ag, and Cu, so that silver, and not gold, carries the largest lattice of the three, while copper is smaller than either by more than 0.3~\r{A}. That ordering is inherited from the bulk metals, and it is what makes the two Cu-bearing pairs the ones that must accommodate a size difference when the species are placed on a common lattice, a point to which the model of Section~\ref{sec:model} returns. In Noblene, the three unlike distances are identical at 2.661~\r{A}, because the equivalence of the three sublattices under that rotation forbids the relaxation from separating them, and this fixes the cell parameter at 4.610~\r{A}. The areal number densities follow the distances in reverse, so that Copperene packs the most atoms per unit area and Noblene, at 16.3~nm$^{-2}$, falls between Silverene and Copperene. In mass, the ordering is set by the atomic weights instead, at 5.04, 2.66, 2.09, and 3.32~mg/m$^{2}$ for Goldene, Silverene, Copperene, and Noblene, so that the alloy sits below gold and above the two lighter metals, as the mean of the three masses on a lattice of intermediate size requires.

With the parent lattice fixed, the space of arrangements it supports is entirely combinatorial, and the first two stages of that workflow are to define that lattice and to enumerate the space exhaustively. Every ordering of Au, Ag, and Cu on the close-packed plane is a derivative superstructure of the one-atom parent cell, so the symmetrically distinct supercells of a given size are generated from the Hermite normal forms of that cell. The distinct decorations of each supercell are obtained by removing the redundancy of the affine symmetry group~\cite{Hart2008, Hart2009, Hart2012}. The number of symmetry-inequivalent orderings at a given cell size then follows from the cycle structure of that group,

\begin{equation}
N_{\mathrm{str}} = \sum_{\mathrm{HNF}} \frac{1}{|G|} \sum_{g} 3^{\,c(g)},
\label{eq:polya}
\end{equation}

\noindent where the outer sum runs over the Hermite normal forms of the given size, the inner sum runs over the elements $g$ of the symmetry group $G$ acting on the sites of each supercell, $c(g)$ is the number of cycles into which $g$ partitions those sites, and the base three counts the species available at every site. Evaluating Eq.~\ref{eq:polya} returns 3, 3, 14, and 39 orderings for cells of one, two, three, and four atoms, the counts shown in Figure~\ref{fig:fig01}(e), and the same expression continues as 72, 237, 465, 1893, 3960, 12672, 25245, and 134083 through twelve atoms per cell, a cumulative 178686. This growth makes exhaustive treatment a decision about where to stop rather than a matter of effort, and we have relaxed the enumeration in full through four atoms per cell, which is the largest size at which every member of the set can be computed individually. Those 59 structures are listed with their compositions and mixing energies in Tables~S1 and~S2 of the Supplementary Information, and their relaxed geometries are shown in Figure~S1.

Each enumerated ordering was then relaxed under a common protocol, the third stage of the workflow, with the cell degrees of freedom restricted to the plane of the layer and the ionic positions and cell shape optimized simultaneously. All of the energetics reported here is referred to a single quantity, the mixing energy
\begin{equation}
E_{\mathrm{mix}} = \frac{E}{N} - \sum_{i \in \{\mathrm{Au,\,Ag,\,Cu}\}} x_{i}\,E_{i}^{\mathrm{pure}},
\label{eq:mix}
\end{equation}
\noindent in which $E$ is the total energy of the relaxed cell, $N$ the number of atoms it contains, $x_{i}$ the fraction of species $i$, and $E_{i}^{\mathrm{pure}}$ the energy per atom of the pure monolayer of that species relaxed on the same lattice with the same parameters. The three pure monolayers are therefore the zero of the scale by construction rather than by measurement, and a negative $E_{\mathrm{mix}}$ means that the alloy is bound against separation into the pure layers on that lattice. Referring the scale to the pure monolayers rather than to the bulk metals keeps the comparison entirely within two dimensions and isolates the alloying chemistry from the energy cost of reducing the coordination, which is common to every structure in the set.

\subsection{Ordering versus composition}\label{sec:ordering}

\begin{figure*}[t]
\centering
\includegraphics[width=0.8\linewidth]{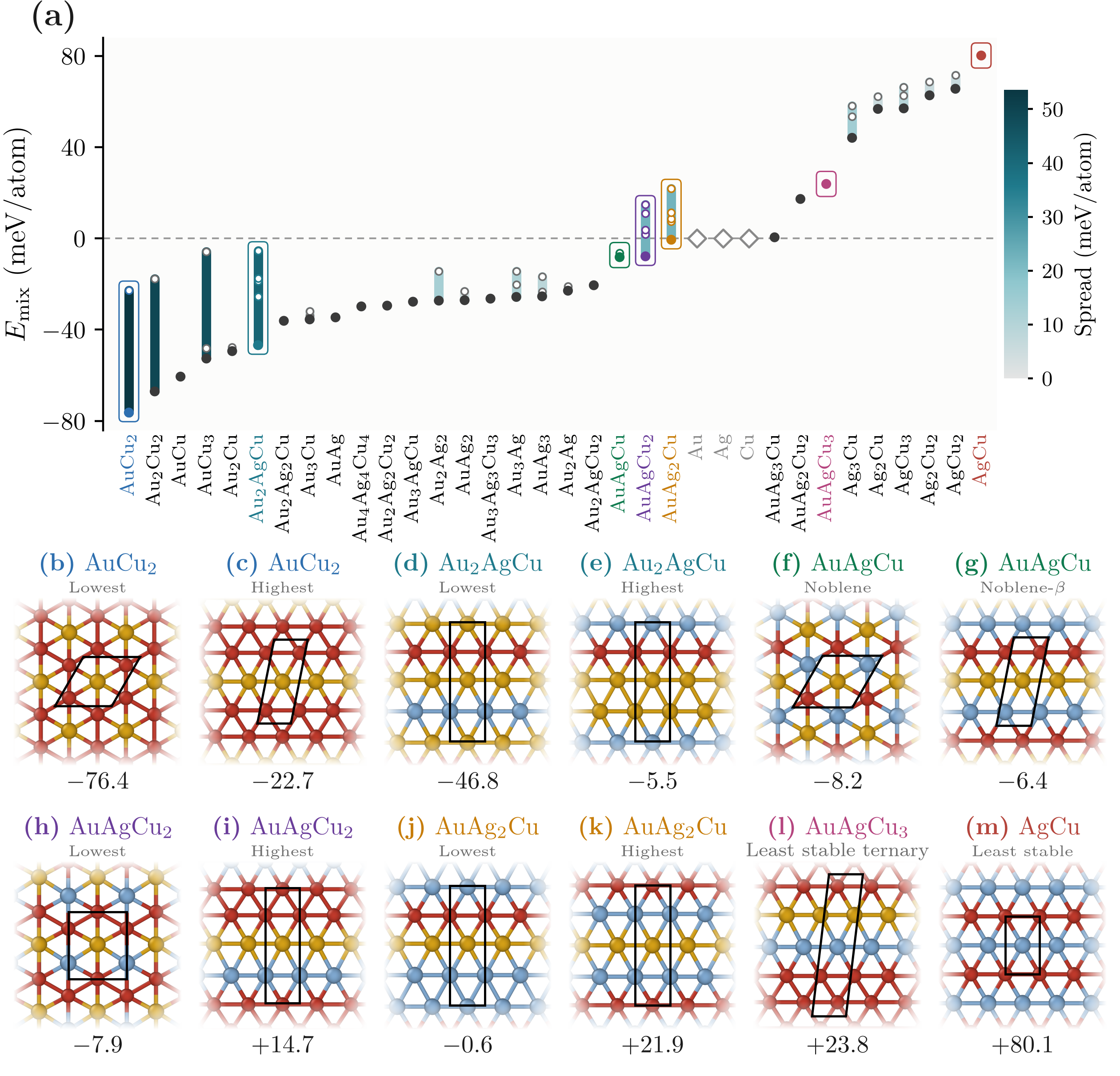}
\caption{Mixing energy of the 68 relaxed structures grouped by cell stoichiometry (a), with each capsule spanning the structures computed at one stoichiometry, colored by the width of that span, and with the three pure monolayers as open reference symbols. Lowest and highest orderings at AuCu$_2$ (b, c) and at Au$_2$AgCu (d, e), the two equimolar orderings Noblene and Noblene-$\beta$ (f, g), the lowest and highest orderings of the two compositions whose sign is set by the arrangement, AuAgCu$_2$ (h, i) and AuAg$_2$Cu (j, k), the least stable ternary AuAgCu$_3$ (l), and the least stable structure of the set AgCu (m). Values below each panel are in meV/atom.}
\label{fig:fig02}
\end{figure*}

A substitutional alloy on a fixed lattice carries two independent degrees of freedom, the proportion of each species and the way those species are distributed over the sites, and the two are conventionally treated as if the first dominated the second. Phase diagrams are drawn against composition, and the ordering degree of freedom is normally collapsed into an average through a mean-field or random-mixing description. Whether that hierarchy survives in a single-atom-thick close-packed layer, in which every atom has six neighbors instead of twelve and no second layer averages the local environment, is the first question addressed here, and Figure~\ref{fig:fig02} collects the evidence that answers it.

The overview in Figure~\ref{fig:fig02}(a) places all 68 relaxed structures on one axis, grouped into the 34 cell stoichiometries they realize, that is, by the formula of the cell rather than by the reduced composition, and ordered from the stoichiometry with the lowest structure found to the one with the highest. Each stoichiometry is drawn as a capsule spanning the structures computed there and colored by the width of that span, and the three pure monolayers appear as open reference symbols at the zero of the scale. The set covers more than 150~meV/atom, running from an ordering of AuCu$_2$ at $-76.4$~meV/atom to an ordering of AgCu at $+80.1$~meV/atom, and the sequence is itself smooth, with neighboring entries differing only slightly in their lowest energies. Two features of the panel are worth noting before examining any of the structures. First, the axis is sorted by gold. Every entry with a negative mixing energy contains gold, the six least stable entries are the six Ag-Cu binaries, which are the only gold-free entries in the set, and no silver appears until the sixth entry, Au$_2$AgCu, the most stable ternary of the set. This ordering already separates the two Cu-bearing pairs. Second, the vertical extent of the capsules is not smooth at all, and in several cases the spread within one stoichiometry exceeds the total variation across many consecutive ones.

Reading Figure~\ref{fig:fig02}(a) across the sequence rather than within it turns that second observation into a number. The median spread of the mixing energy inside a single stoichiometry, taken over the 19 that admit more than one ordering, is 9.3~meV/atom, while the median step between neighboring stoichiometries across the full sequence is 2.5~meV/atom, a ratio of 3.7. Entries realizable in only one arrangement are excluded from the first median because they carry no ordering freedom to measure, and they are the ones at the vertices and at the simplest binary ratios. Merging the stoichiometries that differ only by a common factor, so that the grouping is by reduced composition, raises the ratio to 5.4, and the two groupings are compared in Section~S5 of the Supplementary Information. On either grouping, rearranging a fixed set of atoms moves the energy several times as far as changing which atoms are present, which inverts the hierarchy that composition-based descriptions assume and means that a stoichiometry alone does not identify a material on this lattice. The freedom that a given composition retains to be ordered is itself uneven across the triangle, being smallest at the vertices and along the binary edges and largest in the ternary interior where the equimolar composition lies, as mapped in Figure~S3 of the Supplementary Information.

The panels of Figure~\ref{fig:fig02} take that statement apart composition by composition, and the extreme case is AuCu$_2$, whose two orderings are shown in Figure~\ref{fig:fig02}(b) and (c). The lowest lies at $-76.4$~meV/atom and the highest at $-22.7$~meV/atom, so two arrangements of the same three atoms in cells of the same size are separated by 53.6~meV/atom, more than five times the median spread found within a composition. Both are bound with respect to the pure monolayers, but the first is the most stable structure found anywhere in this work while the second lies well inside the range spanned by the remaining structures, and nothing in the composition distinguishes them.

Silver enters the set at Au$_2$AgCu, whose lowest and highest orderings are shown in Figure~\ref{fig:fig02}(d) and (e). The two span $-46.8$ to $-5.5$~meV/atom, a spread of 41.3~meV/atom at a composition that is already ternary, which shows that the sensitivity to arrangement is not a peculiarity of the binaries and survives the introduction of a third species. The lowest of the two is the most stable ternary computed here, and its bond census contains exactly zero Ag-Cu contacts against one Au-Ag and one Au-Cu bond per atom, an arrangement that anticipates the bond analysis below.

The equimolar composition supports the two orderings of Figure~\ref{fig:fig02}(f) and (g), of which the first is Noblene, at $-8.2$~meV/atom, in which the three species tile the plane so that no atom has a neighbor of its own kind. The second, which we call Noblene-$\beta$, lies at $-6.4$~meV/atom and arranges the same three atoms in alternating rows, so that each species does contact itself along one lattice direction. Both are bound against the pure monolayers. The 1.8~meV/atom that separates them is small on the scale of Figure~\ref{fig:fig02}(a), and it is the difference between an arrangement with no like-species contact and the closest arrangement that has one.

At two compositions the ordering does not merely displace the mixing energy but determines its sign, as the four panels of Figure~\ref{fig:fig02}(h) to (k) show. AuAgCu$_2$ runs from $-7.9$ to $+14.7$~meV/atom between panels (h) and (i), and AuAg$_2$Cu from $-0.6$ to $+21.9$~meV/atom between panels (j) and (k), and the two compositions differ from one another only in whether the third species is copper or silver. The copper-rich one is the more stable at both ends of its range, which is the first indication in the figure that the two Cu-bearing pairs act in opposite directions. Within either composition the same collection of atoms is either bound or unbound against the pure monolayers depending only on where each species sits, so a description that assigns a single mixing energy to a composition would report both as marginal, when in practice they are simultaneously favorable and unfavorable.

The unbound end of the sequence closes the argument, since the least stable ternary of the set is AuAgCu$_3$ at $+23.8$~meV/atom, shown in Figure~\ref{fig:fig02}(l), a five-atom cell drawn from the model-selected part of the set rather than from the exhaustive enumeration, whose own least stable ternary is the AuAg$_2$Cu ordering already shown in Figure~\ref{fig:fig02}(k). What makes the AuAgCu$_3$ arrangement unfavorable is not an excess of Ag-Cu contact, since silver distributes its neighbors evenly there and the Ag-Cu, Au-Cu, and Au-Ag densities are all 0.4 bonds per atom, but rather that the Au-Cu bonds which would stabilize it are exactly as scarce as the Ag-Cu bonds which penalize it, so that the difference between the two densities vanishes while Cu-Cu contacts dominate the census at 1.4 per atom. The least stable structure found anywhere in this work, the AgCu ordering of Figure~\ref{fig:fig02}(m) that closes the top of the range, carries two Ag-Cu bonds per atom against 0.5 each of Ag-Ag and Cu-Cu. That is the largest Ag-Cu density the lattice admits at this composition, and it falls short of the six unlike neighbors per atom that the coordination would allow because the triangular lattice is not bipartite, so that no equimolar binary arrangement on it can avoid like-species contact altogether. None of the three Ag-Cu structures computed here exceeds two. The obstruction is specific to two species, and it is what makes the equimolar ternary examined in Section~\ref{sec:Noblene} the case in which a fully unlike tiling does exist. Read together with Figure~\ref{fig:fig02}(b), where the most stable structure of the set is the one that maximizes Au-Cu contacts, the two extremes of the figure identify the quantity that governs the energetics as the census of bonds rather than the census of atoms.

\begin{figure}[h!]
\centering
\includegraphics[width=0.8\columnwidth]{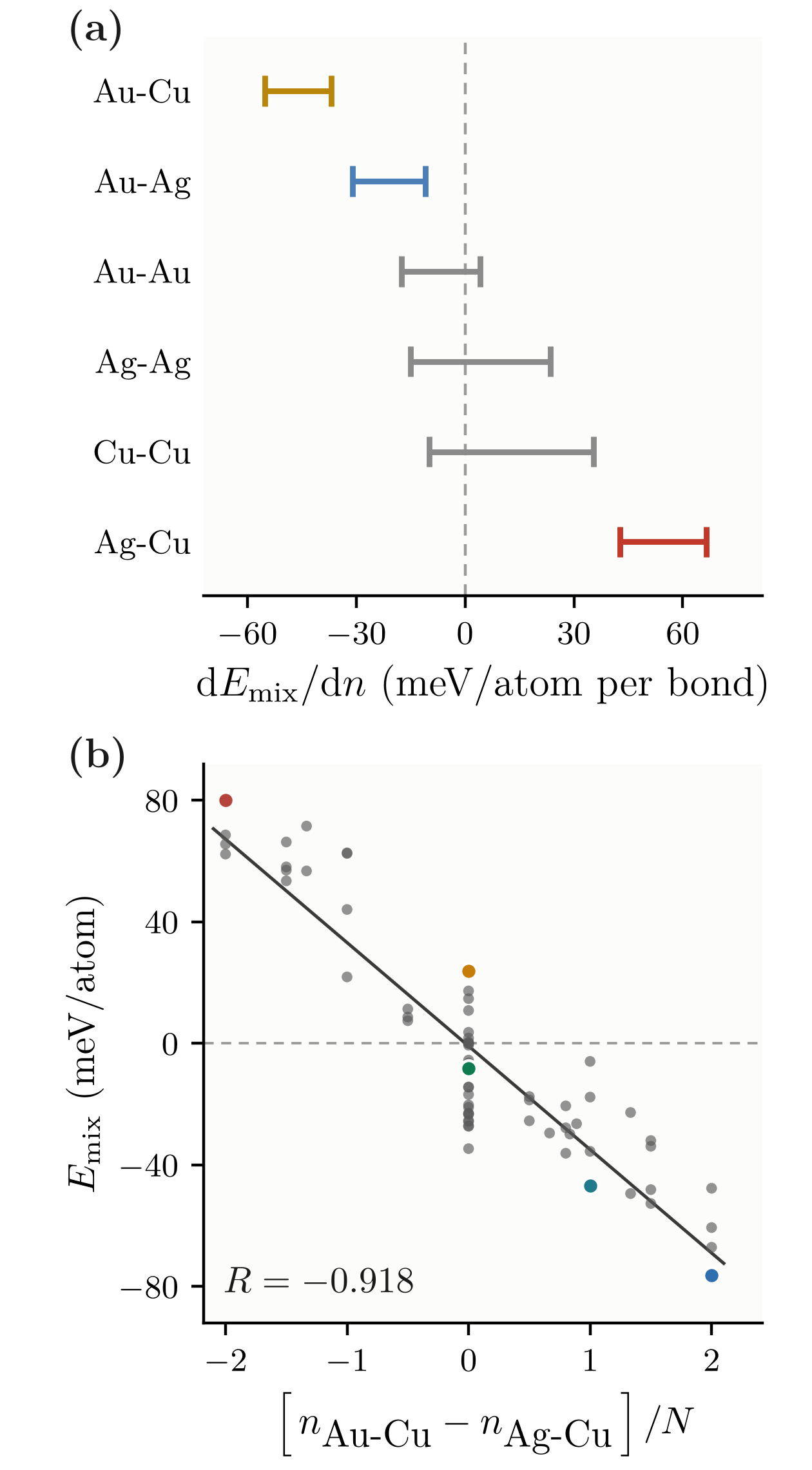}
\caption{Slope of the mixing energy with respect to the density of each of the six bond types, with 95\% confidence intervals, computed over the structures that contain the bond in question (a), and mixing energy against the balance between the Au-Cu and Ag-Cu bond densities, with the linear fit and the corresponding correlation coefficient (b).}
\label{fig:fig03}
\end{figure}

What separates one arrangement from another at fixed composition is therefore not the number of atoms of each species, which is fixed by definition, but the number of bonds of each type. On a triangular lattice each site has six neighbors, and the six pair counts per atom are not independent, because summing the bond ends belonging to each species fixes the three like-type counts once the composition and the three unlike-type counts are specified. The mixing energy can therefore be interrogated against three numbers rather than six, and Figure~\ref{fig:fig03} does exactly that, using bond counts read from the relaxed structures and no fitted parameter of any kind.

Figure~\ref{fig:fig03}(a) gives the derivative of the mixing energy with respect to the density of each bond type, evaluated only over the structures that actually contain that bond, since a structure with no Ag-Cu contacts carries no information about the effect of Ag-Cu contacts. Ag-Cu bonds raise the mixing energy by $55 \pm 12$~meV/atom per bond with a correlation of $R = +0.84$, Au-Cu bonds lower it by $46 \pm 9$~meV/atom per bond with $R = -0.85$, and Au-Ag bonds lower it more weakly, by $21 \pm 10$~meV/atom per bond with $R = -0.56$. The three like-type contacts, Au-Au, Ag-Ag, and Cu-Cu, are not resolved from zero, as expected once lattice identity is accounted for, because they are determined by composition and the unlike-type counts and carry no independent chemistry.

The competition between the two Cu-bearing pairs is therefore what sets the energetics, and Figure~\ref{fig:fig03}(b) reduces the picture to a single coordinate. Plotting the mixing energy against the balance between the Au-Cu and Ag-Cu bond densities per atom yields $R = -0.92$ over the whole set, so one difference of two bond counts accounts for most of the energetics of every structure computed here, pure, binary, and ternary alike. The sign pattern reproduces the bulk alloys, in which Au-Cu orders, Ag-Cu separates, and Ag-Au orders weakly~\cite{Ozolins1998, Cao2007}. To put the magnitudes on a common footing we have computed the bulk mixing enthalpies of the three equimolar binaries in the L1$_0$ structure under the same functional, the same pseudopotentials, and the same convergence criteria used for the monolayers, obtaining $-50.7$~meV/atom for Cu-Au, $-59.8$~meV/atom for Ag-Au, and $+103.3$~meV/atom for Cu-Ag, with the relaxed bulk geometries and the resulting ratios collected in Section~S4 of the Supplementary Information. Comparing these with the two-atom equimolar orderings computed here shows that the reduction to two dimensions does not rescale the three pairs by a common factor. Cu-Au reaches 1.19 times its bulk value, so that the ordering tendency of that contact is stronger in the monolayer than in the crystal, whereas Ag-Au falls to 0.58 and Cu-Ag to 0.78 of their respective values. The contact that dominates the two-dimensional energetics is therefore the one that gains on leaving three dimensions, while the two that weaken are the pair that orders only marginally and the pair that separates. The chemistry of noble-metal alloying therefore survives the reduction to two dimensions with its hierarchy intact and its dominant term reinforced.

Bond counting nevertheless does not exhaust the ordering degree of freedom, since structures whose six bond densities coincide still differ by as much as 29.6~meV/atom, which places the remainder in the geometry of the arrangement rather than in its bond census. A description built on bond counts therefore carries the trend between compositions, and it is on that basis that the next subsection extends the enumeration to the whole ternary triangle.


\subsection{Pair model and landscape}\label{sec:model}

Exhaustive enumeration reaches only the smallest cells, so the compositions open to direct calculation form a sparse and unevenly spaced sample of the ternary triangle, clustered on the simple ratios that three and four atoms can express. Any statement about which compositions the alloy favors therefore requires a description that carries the information contained in the relaxed structures to compositions that no tractable cell can realize, and the relevant question is how few terms such a description requires.

The fourth and fifth stages of the workflow of Figure~\ref{fig:fig01}(f) address that requirement, the first by fitting a model to the relaxed set and the second by carrying it to lattices that direct calculation cannot reach. We have described the mixing energy as a sum of three pair terms, one for each unlike pair, weighted by the number of bonds of that type per atom, together with a single term proportional to the composition-weighted mean square difference of the nearest-neighbor distances of the pure monolayers,
\begin{equation}
E_{\mathrm{mix}}^{\mathrm{model}} = \sum_{i<j} W_{ij}\,\frac{n_{ij}}{N} + K \sum_{i<j} x_{i} x_{j} \left( r_{i} - r_{j} \right)^{2},
\label{eq:model}
\end{equation}
\noindent where $n_{ij}$ counts the nearest-neighbor bonds between species $i$ and $j$ in the cell, $x_{i}$ is the fraction of species $i$, and $r_{i}$ is the relaxed nearest-neighbor distance of the pure monolayer of that species, the values given in Section~\ref{sec:systems}. The first sum is the pair truncation of a cluster expansion on the fixed parent lattice~\cite{Sanchez1984, Connolly1983, deFontaine1994}, and the second carries the elastic cost of size mismatch, which a configurational expansion on a rigid lattice does not represent~\cite{Laks1992}. The three pair terms carry the chemistry of each contact and change when the species are rearranged at fixed composition, whereas the fourth term depends on composition alone and measures the elastic cost of forcing atoms of different size onto a common lattice. Like-species terms are absent because the bond census does not resolve them from zero, so the model has four parameters.

The choice of a size term over higher-order clusters was made by direct comparison rather than by assumption. Three pair parameters alone reproduce the computed set with a mean absolute error of 6.8~meV/atom and a leave-one-out error of 7.2~meV/atom. Adding the seven independent trio terms, which brings the parameter count to ten, improves those figures only to 6.2 and 6.8~meV/atom. Replacing the trios with the single elastic term, for four parameters in total, gives 5.5 and 5.9~meV/atom with $R^{2} = 0.96$, so one composition-dependent parameter accomplishes more than seven configuration-dependent ones. The reason a term of this kind resists absorption into the cluster hierarchy is that its dependence on concentration is not linear in the site occupations, and expansions in the standard basis are not guaranteed to converge to a finite Ising-type form in that situation \cite{Sanchez2010}.

The separation the model enforces between size and chemistry has a direct counterpart in the bulk alloys, where Cu-Au and Cu-Ag carry the same size mismatch and yet one orders while the other separates, a contrast that has been attributed to electronegativity rather than to size \cite{Ozolins1998}, and the same pattern survives the reduction to two dimensions. Taking the nearest-neighbor distances of the pure monolayers as the size scale, the Au-Cu and Ag-Cu mismatches are comparable to one another while Au-Ag is almost degenerate in size, so the elastic term penalizes the two Cu-containing pairs by nearly the same amount and the entire distinction between the ordering Au-Cu contact and the segregating Ag-Cu contact is carried by the pair parameters. Descriptors built on mean square size differences are routinely used to rationalize solid-solution formation in multicomponent alloys \cite{Zhang2008, Guo2011, TodaCaraballo2015}, and the present fit places that quantity on the same footing as the chemical terms instead of treating it as an empirical criterion applied after the fact.

\begin{figure}[h!]
\centering
\includegraphics[width=0.8\columnwidth]{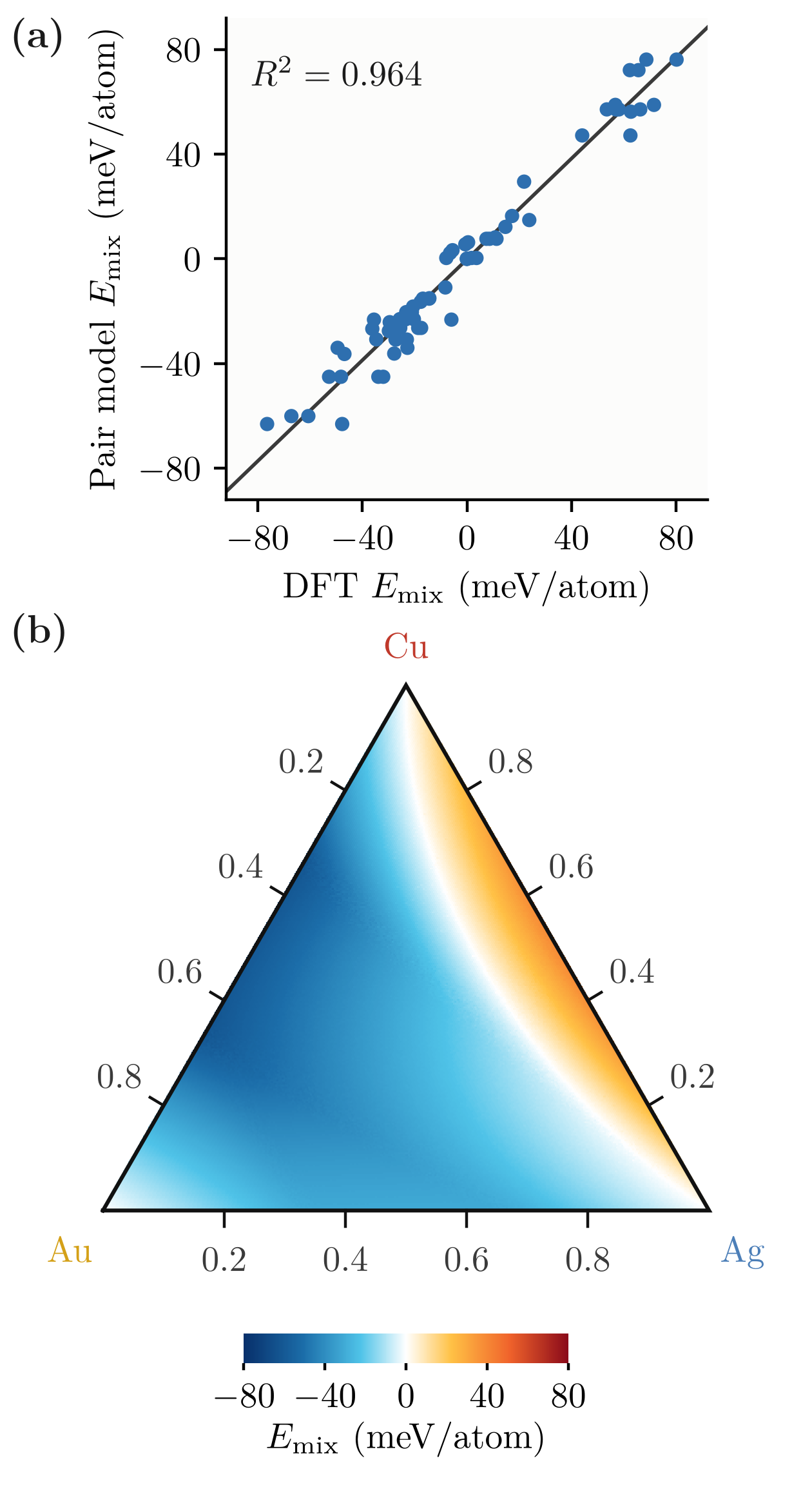}
\caption{Pair model with an elastic size-mismatch term against the computed mixing energies (a) and the modeled mixing energy over the Au-Ag-Cu composition triangle (b).}
\label{fig:fig04}
\end{figure}

The fitted parameters are $W_{\mathrm{Au\text{-}Ag}} = -15.71$, $W_{\mathrm{Au\text{-}Cu}} = -43.66$ and $W_{\mathrm{Ag\text{-}Cu}} = +19.91$~meV/atom, with $K = 1.044$~eV/\AA$^{2}$, and their signs and relative magnitudes reproduce the bond-resolved analysis of the preceding subsection without having been constrained to do so. Figure~\ref{fig:fig04}(a) compares the model against the computed mixing energies across the whole set.

The model has been checked out of sample. The fit was restricted to the 59 exhaustively enumerated structures with $N \leq 4$ and then used to predict the nine larger cells with $N = 5$, 6, 9, and 12, which had entered no part of the fitting procedure. The mean absolute error on those nine is 5.1~meV/atom against 5.6~meV/atom in sample, so the model does not degrade when it leaves the cells that built it. The individual residuals are tabulated in Table~S3 of the Supplementary Information. Extending the fit from $N \leq 3$ to the complete set moves $W_{\mathrm{Au\text{-}Ag}}$ by 3\% and $W_{\mathrm{Au\text{-}Cu}}$ by 6\%, with the pair parameters essentially stationary once the $N = 4$ cells enter, while the residual movement is concentrated in $W_{\mathrm{Ag\text{-}Cu}}$ at 25\% and in $K$, which shifts 15\% from 1.221 to 1.044~eV/\AA$^{2}$. The two quantities that keep moving are the segregating pair and the elastic penalty, both of which are sampled by the fewest structures at small cell size.

With the parameters fixed, the model has been evaluated over 41905 compositions, the complete set of integer compositions of a $24 \times 24$ lattice of 576 sites swept in steps of two atoms, with the ordering at each composition obtained by simulated annealing on that lattice. Figure~\ref{fig:fig04}(b) presents the resulting landscape. The topography is strongly asymmetric and follows the signs of the pair parameters, with a broad valley running along the Au-Cu edge and deepening toward Cu-rich compositions, a ridge of positive mixing energy occupying the Ag-Cu edge, the Au-Ag edge shallow throughout, and the ternary interior intermediate between the two extremes rather than stabilized by the simultaneous presence of all three species. Mixing three noble metals on this lattice therefore does not produce a cooperative gain, which distinguishes the system from the configurational stabilization invoked for multicomponent bulk alloys \cite{Yeh2004, Cantor2004}.

The convex hull of the landscape confirms what the topography suggests, since of its 23 compositions, eighteen lie on the binary edges or at the pure vertices, and the five that do not carry between two and six atoms of a third species in a cell of 576 sites, falling below the envelope spanned by the binaries alone by at most 1.3~meV/atom, which is well inside the accuracy of the model that produced them, so that no balanced ternary composition reaches the hull. The global minimum lies at $x = (0.340, 0.000, 0.660)$ at $-62.9$~meV/atom, which is Au:Cu in a ratio of essentially 1:2 and is already realizable in a cell of three atoms. The deepest balanced ternary composition, Au$_2$AgCu at $x = (0.50, 0.25, 0.25)$, lies above that hull. This result mirrors the bulk ternary, where the phase behavior is dominated by the Cu-Ag miscibility gap with a positive Cu-Ag interaction parameter against negative Cu-Au and Ag-Au \cite{Cao2007, Subramanian1993}, and it indicates that the chemistry responsible for that asymmetry is not modified by the reduction to a single atomic layer. The annealed configurations behind the landscape show the same preference directly, since at compositions in the ternary interior they do not mix but separate into silver-rich regions and Au-Cu regions, which is the real-space counterpart of a hull built entirely from binaries.

The model reproduces the trend from one composition to the next to within a few meV/atom, which is what the landscape of Figure~\ref{fig:fig04}(b) rests on, and the following subsection turns to the structure that the enumeration singles out.


\subsection{Noblene}\label{sec:Noblene}

The sixth and final stage of the workflow of Figure~\ref{fig:fig01}(f) is the verification of a selected structure through its phonon spectrum, its elastic constants, its band structure, and the effect of spin-orbit coupling (SOC), and the structure we carry through it is Noblene, the only arrangement among the 68 relaxed structures in which every atom is surrounded exclusively by unlike species, so that it is the most thoroughly mixed configuration the ternary admits on this lattice and the one that expresses in a single cell the alloying chemistry that the rest of this work has been measuring.

That singularity is not an accident of the enumeration. An arrangement in which no atom has a neighbor of its own kind is a proper three-coloring of the triangular lattice, and because every elementary triangle must carry all three species, assigning one triangle fixes the whole lattice. The arrangement is therefore unique up to which species occupies which sublattice, and it exists only at the equimolar composition. Noblene carries exactly one Au-Ag, one Au-Cu, and one Ag-Cu bond per atom, with Au-Au, Ag-Ag, and Cu-Cu all identically zero. It is reached by ordering rather than by disorder, which sets it apart from the statistical mixing that defines multicomponent solid solutions \cite{Yeh2004, Cantor2004} and from the structures constructed to imitate randomness in a small cell \cite{Zunger1990}.

Its mixing energy is $-8.2$~meV/atom, so the arrangement is bound with respect to the three pure monolayers already in the three-atom cell, which is the smallest cell of this lattice that can carry an equimolar ternary at all. Between the two orderings that this cell admits, Noblene is the more stable, ahead of Noblene-$\beta$ at $-6.4$~meV/atom, whose cell does not retain the hexagonal angle and relaxes to $\gamma = 120.6^{\circ}$, and the margin between the two widens from 1.8 to 3.1~meV/atom when SOC is included, so the ordering preference is reinforced rather than removed by the relativistic terms that matter most for gold \cite{DalCorso2005}.

The equimolar composition also supports larger orderings, and those reach lower energies, between $-26$ and $-30$~meV/atom for the six-, nine-, and twelve-atom cells computed here. Figure~S2 of the Supplementary Information shows that they are not mixed alloys at all but interfaces, each segregating the three species into stripes of like atoms so that the Au-Cu and Au-Ag contacts are maximized along the boundaries while the Ag-Cu contact that the bond analysis of Section~\ref{sec:ordering} identifies as the costly one is minimized. The same preference governs the composition landscape of Section~\ref{sec:model}, whose convex hull no balanced ternary composition reaches. Energy at this composition is therefore obtained by separating rather than by mixing, and Noblene sits at the opposite extreme of that trade.

\begin{figure*}[t]
\centering
\includegraphics[width=0.8\linewidth]{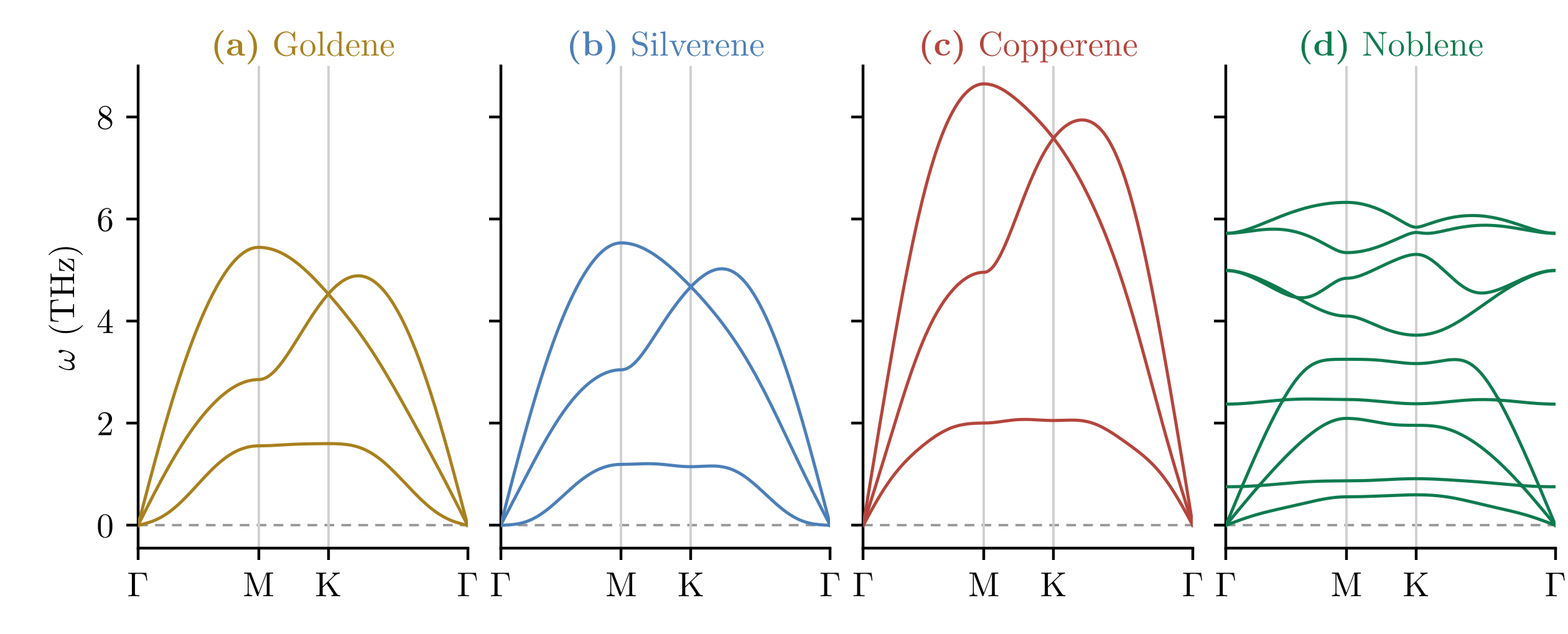}
\caption{Phonon dispersions along $\Gamma$-M-K-$\Gamma$ for Goldene (a), Silverene (b), Copperene (c), and Noblene (d).}
\label{fig:fig05}
\end{figure*}

Thermodynamic distance from a hull built entirely from binaries is a weak criterion for a monolayer that would be prepared by exfoliation rather than by equilibrium solidification, and the operative question is whether the lattice is dynamically stable. Figure~\ref{fig:fig05} presents the phonon dispersions of Goldene, Silverene, Copperene, and Noblene.

All four systems are dynamically stable, with every branch real along $\Gamma$-M-K-$\Gamma$ and the acoustic branches emerging from $\Gamma$ with the linear and quadratic dispersions expected of a free-standing 2D layer. The upper edge of the spectrum orders as the masses require, with 5.45~THz for Goldene, 5.53~THz for Silverene, and 8.65~THz for Copperene, the lightest of the three, while Noblene closes at 6.33~THz, intermediate between the heavy and the light constituents as the mixed masses of its three-atom cell would suggest.

\begin{figure*}[t]
\centering
\includegraphics[width=0.8\linewidth]{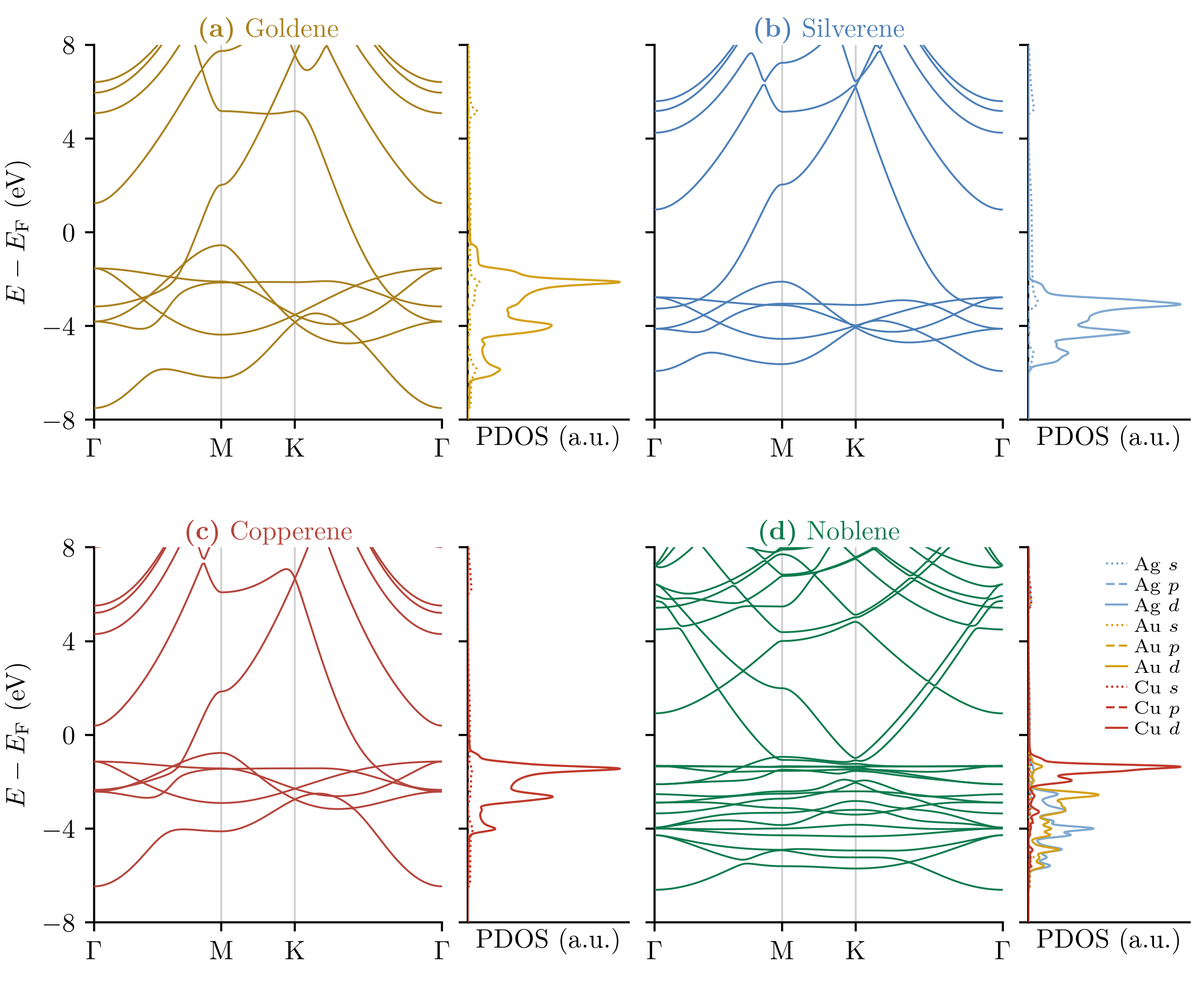}
\caption{Band structures along $\Gamma$-M-K-$\Gamma$ and species- and angular-momentum-resolved PDOS per atom for Goldene (a), Silverene (b), Copperene (c), and Noblene (d). Energies are referred to the Fermi level.}
\label{fig:fig06}
\end{figure*}

The force constants that set those frequencies are themselves a property of the electronic ground state, so the same substitution can be followed one level further down, into the states from which they derive. Because the three constituents carry $d$ manifolds centered at different energies, replacing a single species by three inequivalent ones redistributes the occupied weight across the valence region, and whether that redistribution reaches the Fermi level is what decides the metallic character of the ternary. The three pure monolayers have been characterized as metals in previous first-principles work~\cite{dosSantos2025, Wang2025, Xu2026}, and they serve here as the reference against which the ternary is read. Figure~\ref{fig:fig06} presents the band structures and the species-resolved projected densities of states (PDOS) of the four systems.

Noblene is metallic, with bands crossing the Fermi level, a finite PDOS there, and no gap and no pseudogap at that energy. The density of states at the Fermi level is 0.214~states/eV per atom, against 0.200, 0.213, and 0.232 for Goldene, Silverene, and Copperene, whose composition average is 0.215, so the alloy reproduces the average of its constituents to within 0.001~states/eV per atom, and the maximally mixed arrangement neither depletes nor enhances the states available for conduction. Resolved by site, the three atoms keep the ordering of the elements they came from, with 0.248 on Cu, 0.211 on Ag, and 0.184 on Au, in the same sequence as Copperene, Silverene, and Goldene. Every site carries an $s$ and $p$ contribution larger than its $d$ contribution at $E_{\mathrm{F}}$, so the dispersive free-electron-like band of the pure monolayers survives the substitution.

The three species remain distinguishable in their $d$ manifolds, whose centers lie at $-3.33$, $-3.79$, and $-1.75$~eV for Au, Ag, and Cu in Noblene against $-3.14$, $-3.72$, and $-2.12$~eV in the corresponding pure layers, so the elemental ordering with Cu highest and Ag lowest is preserved. The largest shift, 0.37~eV upward for Cu, remains small beside the 2.04~eV that separates the Cu manifold from the Ag one. The bandwidths report the change in neighborhood more directly than the centers do, since no atom in Noblene has a neighbor of its own species, so the $d$-$d$ pathway between like atoms is cut and the Au manifold narrows from 1.45 to 1.25~eV. The Ag manifold instead broadens from 0.94 to 1.17~eV, which is what the energies anticipate, since Au and Ag lie only 0.46~eV apart in Noblene and hybridize with one another, while Cu sits between 1.6 and 2.0~eV above both and is left essentially untouched at 0.93~eV against 0.94~eV in Copperene. The three metals therefore neither merge into a common band nor lose their separate $d$ signatures, and the pair closest in energy is the pair that mixes, which is the electronic counterpart of the bond census that makes Noblene singular.

The same bonds whose census governs the mixing energy also set the curvature of the energy under strain, and that curvature is what the elastic constants of Table~\ref{tab:elastic} measure. In a 2D crystal with hexagonal symmetry, the in-plane elastic response is exactly isotropic, so the Young's modulus and the Poisson's ratio carry no angular dependence and a polar representation would reduce to four circles. The isotropy is not assumed but recovered by the calculation, since the identity $C_{66} = (C_{11} - C_{12})/2$ holds to within $0.0$--$0.2$~N/m in all four systems and $|C_{11} - C_{22}|$ stays below 0.4~N/m, which also provides an internal consistency check on the finite-strain protocol.

\begin{table}[t]
\centering
\caption{Elastic constants, Young's modulus, and Poisson's ratio of the three
pure monolayers and Noblene, in N/m, except for the dimensionless Poisson
ratio.}
\label{tab:elastic}
\begin{tabular}{lccccc}
\toprule
System & $C_{11}$ & $C_{12}$ & $C_{66}$ & $Y$ & $\nu$ \\
\midrule
Goldene   & 96.7 & 43.2 & 26.7 & 77.3 & 0.448 \\
Silverene & 54.1 & 23.1 & 15.5 & 44.2 & 0.429 \\
Copperene & 88.2 & 30.8 & 28.8 & 77.5 & 0.349 \\
Noblene   & 69.8 & 35.5 & 17.3 & 51.8 & 0.507 \\
\bottomrule
\end{tabular}
\end{table}

All four systems satisfy the Born criteria for a hexagonal 2D lattice \cite{Mouhat2014}, and Noblene departs from the behavior that a simple interpolation between its constituents would predict. Its Young's modulus of 51.8~N/m falls below that of Goldene and Copperene, which are close to one another at 77.3 and 77.5~N/m, and approaches the value of Silverene, the softest of the four at 44.2~N/m, so the stiffness of the alloy is governed by its most compliant constituent rather than by the average of the three. Its Poisson's ratio of 0.507 is the highest of the set, so the transverse contraction accompanying a uniaxial extension is the largest of the four, and since $\nu = C_{12}/C_{11}$ for an isotropic 2D crystal, the stability requirement $C_{11} > |C_{12}|$ places the admissible range at $|\nu| < 1$ rather than at the bound of one half that applies to three-dimensional isotropic solids. That ratio is high in Noblene because $C_{12}$ is the second largest of the set while $C_{11}$ falls well below the values of Goldene and Copperene, so the transverse coupling is retained while the longitudinal stiffness is not, and it is that combination rather than an unusually soft lattice that produces it. The moduli of the pure monolayers are consistent with those reported previously for the same three layers~\cite{dosSantos2025}, with Goldene and Silverene reproduced to within a few percent and Copperene stiffer here, within the spread produced by different cell and strain protocols at the same level of theory.

The route to a material of this kind is already established for one of its constituents, since the chemical exfoliation of an Au-intercalated MAX phase that yielded Goldene \cite{Kashiwaya2024} has since been extended to the trilayer through the corresponding Ti$_4$Au$_3$C$_3$ precursor \cite{Shi2025}. A separate confinement route has delivered a family of metals at the angstrom thickness limit \cite{Zhao2025}. Since Noblene is an ordered arrangement on that same lattice, the natural target is a MAX-phase precursor whose intercalated layer is already the ordered ternary, so that the arrangement is inherited from the parent crystal rather than established after exfoliation \cite{Kashiwaya2025}.

\section{Conclusions}\label{sec:conclusions}

In summary, we have combined an exhaustive enumeration of the derivative superstructures of the Goldene lattice with DFT to establish how ordered Au-Ag-Cu monolayer alloys distribute their energy. Across 68 relaxed structures spanning 34 cell stoichiometries, the arrangement of the atoms rather than their proportion controls the mixing energy, with the median spread within a stoichiometry reaching 3.7 times the median step between neighboring ones and with two compositions whose sign is determined by the ordering alone. The energetics arises from a competition between Au-Cu bonds, which lower the energy, and Ag-Cu bonds, which raise it, and the difference between those two bond densities correlates with the mixing energy at $R = -0.92$ across the whole set. Four parameters suffice to carry that chemistry beyond the enumerated cells, since three first-neighbor pair terms and one elastic size-mismatch term reproduce the computed energies with a mean absolute error of 5.5~meV/atom, predict the larger cells they never saw at 5.1~meV/atom, and extend over 41905 compositions to a convex hull that no balanced ternary composition reaches. Notably, Noblene, the equimolar ordering in which the three species occupy the three sublattices of the close-packed plane, is the only arrangement the lattice admits with no like-species contact anywhere in it, and it is dynamically stable, metallic, and elastically isotropic in plane.

These results place ordering rather than composition at the center of alloy design on single-atom-thick close-packed lattices, and they suggest that the search for 2D alloys is better organized around arrangements at fixed stoichiometry than around composition sweeps alone. The natural continuation is the finite-temperature problem, in which configurational entropy competes with the ordering energies measured here, and the low ordering temperatures established for the bulk noble-metal binaries suggest that the competition is decided within an experimentally accessible window. Noblene is the arrangement that an exfoliation route departing from a mixed noble-metal MAX phase would most plausibly target, and it represents the ordered ternary monolayers well precisely because of that uniqueness.


\section{Methods}\label{sec:methods}

The parent lattice was inherited from Goldene, a single close-packed atomic plane of a face-centered cubic noble metal, which the atlas of elemental 2D metals identifies as the most stable simple 2D geometry in this family \cite{Nevalaita2018}. Enumeration of the derivative superstructures, described in Section~\ref{sec:systems}, follows the Hart-Forcade construction \cite{Hart2008, Hart2009, Hart2012}, and the count at each cell size was verified independently by evaluating Eq.~\ref{eq:polya} over the same group. A preliminary fit of the pair model to the 59 exhaustively enumerated structures was then used to select nine further structures with $N = 5$, 6, 9, and 12, which were relaxed under the same protocol and added to the set. Structure generation and manipulation used the atomic simulation environment \cite{Larsen2017}.

All calculations were performed within DFT \cite{Hohenberg1964, Kohn1965} as implemented in the Quantum ESPRESSO distribution \cite{Giannozzi2009, Giannozzi2017, Giannozzi2020}, with exchange and correlation treated in the generalized gradient approximation of Perdew, Burke, and Ernzerhof \cite{Perdew1996}. Optimized norm-conserving Vanderbilt pseudopotentials \cite{Hamann2013} from the PseudoDojo table \cite{vanSetten2018}, whose transferability for this family has been assessed against all-electron references \cite{Lejaeghere2016}, described the electron-ion interaction. Plane waves were included to 100~Ry, with a charge-density cutoff of 400~Ry. Brillouin-zone sampling used Monkhorst-Pack grids \cite{Monkhorst1976} of $55 \times 55 \times 1$ for the one-atom cells, with the mesh of every other cell scaled with the cell size so that the product of the mesh and the in-plane lattice parameter is held at approximately 150~\AA{}, and with the noninteger occupations near the Fermi level treated by the cold smearing scheme of Marzari and Vanderbilt \cite{Marzari1999} with a width of 0.02~Ry. Periodic images along the direction normal to the layer were separated by 15~\AA{} of vacuum. Self-consistency was converged to $10^{-10}$~Ry, and the structural optimizations to $10^{-5}$~Ry in the total energy and $10^{-4}$~Ry/bohr in the forces.

Relaxations described in Section~\ref{sec:systems} used the variable-cell algorithm, with the vacuum separation preserved by restricting the cell degrees of freedom to the plane. Since a variable-cell run terminates on the basis set of the initial cell, each relaxation was followed by a fresh self-consistent calculation at the relaxed geometry, whose energy enters every quantity reported here. Pure monolayers were relaxed under the identical protocol and provide the reference energies of Eq.~\ref{eq:mix}, with their relaxed nearest-neighbor distances given in Section~\ref{sec:systems}. All quantities are reported per atom so that cells of different sizes can be compared directly.

The hexagonal geometry of the pure monolayers is a genuine minimum rather than a consequence of symmetry locked in a one-atom cell, as verified by relaxing a distorted rectangular two-atom cell for each metal with symmetry detection disabled, all three of which returned to the hexagonal geometry within 7~$\mu$eV/atom of the symmetry-constrained result.

Bulk mixing enthalpies used for comparison in Section~\ref{sec:ordering} were obtained under the same protocol, with the elemental metals relaxed in the face-centered cubic structure and the equimolar binaries in the L1$_0$ structure, and the enthalpy of each binary taken as the difference between its energy per atom and the mean of the energies per atom of its two constituents. Relaxed geometries and the resulting values are collected in Section~S4 of the Supplementary Information.

The four parameters of Eq.~\ref{eq:model} were obtained by linear least squares on the 68 relaxed structures. Model selection among the candidate term sets, namely pairs alone, pairs with the seven independent trios, and pairs with the elastic term, used leave-one-out cross-validation over the full set \cite{Hart2005, vandeWalle2002a, Angqvist2019}. Extrapolation beyond the exhaustively enumerated range was validated by refitting on the 59 structures with $N \leq 4$ alone and predicting the nine structures with $N = 5$, 6, 9, and 12 as a held-out test set.

The composition landscape was obtained by evaluating Eq.~\ref{eq:model} on a triangular lattice of $L = 24$ (576 sites) over the 41905 integer compositions accessible on that lattice in steps of two atoms, a count equal to $\binom{290}{2}$. At each composition, the site occupations were optimized by simulated annealing \cite{Kirkpatrick1983} with Metropolis acceptance \cite{Metropolis1953}, proposing exchanges of pairs of unlike sites under periodic boundary conditions, with a geometric temperature schedule from 400 to 1~meV over $10^{6}$ steps per composition. Its convex hull was constructed over the two independent composition variables.

Phonon dispersions were computed within density functional perturbation theory (DFPT) \cite{Baroni2001}. For Noblene, the dynamical matrices were obtained on a $3 \times 3$ $q$ mesh commensurate with the $33 \times 33$ $k$ mesh of the electronic calculation, and for the pure monolayers a $q$ mesh of $6 \times 6$ against electronic meshes of $55 \times 55$, $54 \times 54$, and $62 \times 62$ was used. Force constants were transformed to real space with the acoustic sum rule imposed. Since the M point is not contained in the $3 \times 3$ mesh, the dynamical matrix there was also computed directly by DFPT for Noblene, and the lowest branch returns $+14.7$~cm$^{-1}$, in agreement with the interpolated value \cite{Carrete2016}.

Band structures were computed along $\Gamma$-M-K-$\Gamma$ from a non-self-consistent calculation on the converged charge density, and the PDOS were resolved by species and by angular momentum and normalized per atom so that cells containing one and three atoms are displayed on the same scale. Centers and widths of the $d$ manifolds are the first and second moments of the $d$-projected density of states, taken in a window from $-10$ to $+3$~eV about the Fermi level, which contains the whole $d$ manifold of the three metals and excludes the semicore states carried by the pseudopotentials. Spin-orbit coupling was included through fully relativistic pseudopotentials in noncollinear calculations \cite{DalCorso2005} for the comparison between the two equimolar orderings.

Elastic constants were obtained from the energy response to finite in-plane strains, with the energy referred to the layer area rather than the cell volume, which places the constants in N/m and removes any dependence on the vacuum thickness. Four deformation modes were applied to the relaxed cell, namely uniaxial strain along each of the two in-plane lattice directions, a biaxial strain, and a pure shear, each isolating one combination of $C_{11}$, $C_{22}$, $C_{12}$, and $C_{66}$. At every strained configuration, the cell was held fixed, the internal coordinates were relaxed, and the electronic $k$ mesh was kept at the one converged for the undeformed cell. Each constant was extracted from a parabolic fit of the energy density over strains of $\pm3\%$ in seven steps. Neither $C_{11} = C_{22}$ nor $C_{66} = (C_{11} - C_{12})/2$ is imposed at any point of this protocol. The in-plane Young's modulus and Poisson's ratio follow from the standard 2D relations \cite{Andrew2012}, and mechanical stability was verified against the Born criteria \cite{Mouhat2014}.

\backmatter

\bmhead{Acknowledgments}

M.L.P.J. acknowledges financial support from FAPDF (grant 00193-00001807/2023-16), CNPq (grants 444921/2024-9 and 308222/2025-3), and CAPES (grant 88887.005164/2024-00).

\section*{Declarations}

\bmhead{ORCID}
Marcelo Lopes Pereira Junior, https://orcid.org/0000-0001-9058-510X.

\bmhead{Competing interests}
The author declares no competing interests.

\bmhead{Data availability}
The relaxed geometries of all 68 monolayers are provided as crystallographic information files, described in Section~S6 of the Supplementary Information. The computed energies and the fitted model parameters are available from the author on reasonable request.

\bmhead{Code availability}
The enumeration, fitting, and annealing scripts are available from the author on reasonable request.

\bibliography{references}

\end{document}